%% file: main.tex
\documentclass[usenatbib]{mnras}
\usepackage{aas_macros}
\usepackage{natbib}
\usepackage[utf8]{inputenc}
\usepackage{graphicx}
\usepackage{amsmath}
\usepackage{amssymb}
\usepackage[dvipsnames]{xcolor}
\usepackage{hyperref}
\hypersetup{colorlinks=true,allcolors=teal}
\usepackage{microtype}
\usepackage[normalem]{ulem}
\usepackage{caption}
\usepackage{subcaption}
\usepackage{tabularx}
\newcommand{\Mh}{\ensuremath{h^{-1}M_{\odot}}}
\newcommand{\Mpch}{\ensuremath{h^{-1}{\rm Mpc}}}
\newcommand{\kpch}{\ensuremath{h^{-1}{\rm kpc}}}

\newcommand{\avg}[1]{\ensuremath{\left\langle \,#1\, \right\rangle}}

\newcommand{\eqn}[1]{equation~\eqref{#1}}

\newcommand{\be}{\begin{equation}}
\newcommand{\ee}{\end{equation}}

\title[Halo properties from halo environment]{Mock halo catalogs: assigning unresolved halo properties using correlations with local halo environment}
\author[Ramakrishnan et al.]{
Sujatha Ramakrishnan$^{1}$\thanks{E-mail: rsujatha@iucaa.in}, 
Aseem Paranjape$^{1}$\thanks{E-mail: aseem@iucaa.in}
\& Ravi K. Sheth$^{2,3}$\thanks{E-mail: shethrk@physics.upenn.edu}
\\  
 $^1$ Inter-University Centre for Astronomy \& Astrophysics,
      Ganeshkhind, Post Bag 4, Pune 411007, India\\
 $^2$ Center for Particle Cosmology, University of Pennsylvania, 209 S. 33rd St., Philadelphia, PA 19104, USA\\
 $^3$ The Abdus Salam International Center for Theoretical Physics, Strada Costiera, 11, Trieste 34151, Italy}

\date{draft}

\begin{document}
\label{firstpage}
\pagerange{\pageref{firstpage}--\pageref{lastpage}}
\maketitle

\begin{abstract}
Large-scale sky surveys require companion large volume simulated mock catalogs. To ensure precision cosmology studies are unbiased, the correlations in these mocks between galaxy properties and their large-scale environments must be realistic.  Since galaxies are embedded in dark matter haloes, an important first step is to include such correlations -- sometimes called assembly bias -- for dark matter haloes.  However, galaxy properties correlate with smaller scale physics in haloes which large simulations struggle to resolve.  We describe an algorithm which addresses and largely mitigates this problem.  Our algorithm exploits the fact that halo assembly bias is unchanged as long as correlations between halo property $c$ and the intermediate-scale tidal environment $\alpha$ are preserved. Therefore, knowledge of $\alpha$ is sufficient to assign small-scale, otherwise unresolved properties to a halo in a way which preserves its large-scale assembly bias accurately.  We demonstrate this explicitly for halo internal properties like formation history (concentration $c_{\rm 200b}$), shape $c/a$, dynamics $c_{v}/a_{v}$, velocity anisotropy $\beta$ and angular momentum (spin $\lambda$).  Our algorithm increases a simulation's reach in halo mass and number density by an order of magnitude, with improvements in the bias signal as large as 45\% for 30-particle haloes, thus significantly reducing the cost of mocks for future weak lensing and redshift space distortion studies.  
\end{abstract}

\section{Introduction}
Future large-volume surveys need thousands of realisations of simulated data of comparable size to serve as test-beds on which to design observables and provide error estimates \citep{2018ApJS..234...36M}. State-of-the-art hydrodynamical simulations are computationally expensive and cannot meet the demands of these surveys in terms of volume or number of realisations without compromising on the resolution.  This has led to dark matter halo-based approaches which simulate the dark matter component only, and then `paint' galaxies onto the simulated dark matter haloes.  These are of two types.  In semi-analytic galaxy formation models, galaxy properties are determined by modeling a range of (baryonic) physical processes that are not present in the dark matter only simulation, on a halo-by-halo basis \citep[see][for a recent review]{2015ARA&A..53...51S}.
These seek to reproduce a wide variety of different observables, but because the underlying physics is uncertain, they are not always guaranteed to reproduce the properties of a given dataset.  Moreover, they are rather computationally expensive. 
The other `Halo Model' approach is more empirically driven, and uses constraints derived from the measured abundance and clustering in a survey to determine how galaxies populate haloes.  The simplest of these assume that the galaxy population in a halo depends on halo mass alone \citep[see, e.g.,][and references therein]{ss09,2016MNRAS.457.4360Z}.  

However, even at fixed mass, stochasticity in halo assembly results in a wide range of structural properties.  These include the halo density profile, shape, angular momentum, velocity structure, etc. 
The assembly history of a halo also affects the formation of galaxies in it, so the mix of galaxies in a halo correlates with many of these other factors as well.  Since haloes having different internal properties cluster differently \citep{st04,wechsler+06,2007MNRAS.374.1303C,2010ApJ...708..469F}, the question of how to incorporate these `assembly bias' effects into mock galaxy catalogs has been the focus of many studies \citep{hw13,masaki+13,pkhp15,2016MNRAS.460.2552H,2019MNRAS.490.2718D,2020PhRvD.102h3520S,2020MNRAS.495.5040H,2020arXiv200705545X,caz21}.  

Our approach to this problem is as follows:  Since the formation history of a halo correlates with its structure, by coupling galaxy properties to this structure, one can account for those aspects of galaxy assembly bias which are directly inherited from halo structure or formation history.  Indeed, previous work has shown that correlations in the mock galaxy population which are inherited from halo mass alone are able to reproduce many observed correlations between real galaxies and their environment \citep{phs18b,azpm19}.  So, incorporating other halo structural parameters should result in even more realistic mocks.  Unfortunately, reliable estimates of many of the structural parameters require that the halo be sampled by many particles.  At the low mass end, one needs approximately $10\times$ more particles per halo than are needed to estimate its mass.  As a result, assembly bias pushes even the `paint galaxies into haloes' approach up against the wall of resolution.  It is this problem which has motivated our study. Since this problem has to do with halo rather than galaxy properties, our focus in this work will be on halo catalogs rather than the prescriptions for painting galaxies into them.

Recent work has shown that the correlations between halo internal properties and the matter distribution on large scales can be factorized as arising from two distinct correlations: one between the large-scale halo bias and $\alpha$ -- a suitably defined measure of the tidal environment on an intermediate scale (\citealp{2018MNRAS.476.3631P}) -- and the other between $\alpha$ and internal properties on smaller scales (\citealp{2019MNRAS.489.2977R}; see also \citealp{2008ApJ...687...12D,2009MNRAS.398.1742H,2017MNRAS.469..594B}).  
As the required resolution for estimating $\alpha$ reliably is not more stringent than for estimating halo mass, $\alpha$ can be measured easily even when internal properties of a halo cannot (we show this explicitly below).  Our goal is to leverage knowledge of halo mass and halo-centric $\alpha$ to make realistic `assembly biased' mocks down to the mass scale at which haloes are sufficiently well-resolved.  In effect, we use the tidal environment as a lens for increasing the effective resolution of a simulation, thus increasing its effective dynamic range by an order of magnitude.

The paper is organised as follows. Section~\ref{sec:simulation} describes the simulations and halo properties used in this work.  Section~\ref{sec:propsfromenv} first shows that the assembly bias signal is unchanged if one shuffles internal halo properties around, between haloes of the same mass {\em and} $\alpha$.  This is an explicit demonstration of why $\alpha$ can be used to improve the effective resolution of a simulation. It then provides fitting functions for the probability distribution of halo properties as a function of halo mass and $\alpha$.  
Section~\ref{sec:application} uses these to generate mock halo catalogs from low-resolution simulations, and demonstrates explicitly that they have the same assembly bias as higher resolution simulations.  We summarise in Section~\ref{sec:summary}. The Appendices provide some of the technical analyses relevant to the main text.

\section{Simulations and Halo properties}
\label{sec:simulation}

Here we use three volumes of $N$-body simulations:  2 realisations of $150\Mpch$, 10 realisations of $300\Mpch$ and 3 realisations of $600\Mpch$, each of which evolve $1024^3$  particles of collisionless  CDM using the tree-PM code \textsc{gadget-2} \citep{2005MNRAS.364.1105S}\footnote{ \url{http://www.mpa-garching.mpg.de/gadget/}}  with a $2048^3$ PM grid. The force resolution in the increasing order of simulation volume is $\epsilon=4.9,9.8$ and $19.6 \kpch$.  Due to the increasing particle mass $m_p =2.4 \times 10^8 \Mh, 1.93 \times 10^{9} \Mh$ and $1.54\times 10^{10} \Mh$, we henceforth refer to these three sets of simulations as the \emph{high-}, \emph{medium-} and \emph{low-resolution} simulations, respectively. 

All the simulations have the same spatially flat $\Lambda$CDM cosmology with the following cosmological parameters: total matter density parameter $\Omega_m=0.276$, baryonic matter density $\Omega_{b}=0.045$, Hubble constant $H_{0}=100 h { \rm kms^{-1}{\rm Mpc^{-1}}}$ with $h=0.7$, primordial scalar spectral index $n_{s}=0.961$ and r.m.s linear fluctuations in spheres of radius $8 \Mpch$, $\sigma_{8}=0.811$, with transfer function generated by the code \textsc{camb} \citep{2000ApJ...538..473L}.\footnote{\url{http://camb.info/}} Initial conditions were generated with the code \textsc{music} \citep{2011MNRAS.415.2101H}\footnote{\url{ https://www-n.oca.eu/ohahn/MUSIC/}} using $2^{\rm nd}$-order Lagrangian perturbation theory, at starting redshifts $z_{\rm in}=99,49,99$ for the 150, 300 and 600$\Mpch$ simulation boxes respectively.
We focus on results at $z=0$ in this work.

\subsection{Halo properties from a halo finder}
\label{subsec:haloprops}
We identified haloes using the halo finder \textsc{rockstar} \citep{2013ApJ...762..109B},\footnote{ \url{https://bitbucket.org/gfcstanford/rockstar}} which is based on a 6-D phase space Friends-of-Friends algorithm. To prevent contamination from substructure as well as any spurious objects, we discard sub-haloes and also apply a virial ratio cut-off $2T/|U|\leq 2$ \citep{2007MNRAS.376..215B}. For our main analysis and calibrations, we use well-resolved haloes with $>500$ particles (see Appendix~\ref{app:resoln} for a resolution study). 

Throughout, we quote halo masses $m$ using the $M_{\rm 200b}$ definition, which is the mass enclosed inside the radius $R_{\rm 200b}$ at which the enclosed density is 200 times the background density. $R_{\rm 200b}$ is also used in measuring the local tidal environment of haloes, as discussed later. Here we list the halo properties used in this work, some of which are directly ouput by \textsc{rockstar}, while for a few others we modified the code to output them.

\subsubsection{Halo Concentration}
The halo concentration is defined as
\begin{equation}
 c_{\rm 200b} = R_{\rm 200b}/r_{s},
\end{equation}
where $r_s$ is the scale radius of the NFW profile \citep{1997ApJ...490..493N} fitted to each halo by \textsc{rockstar}. The halo concentration is the most common assembly bias variable discussed in the literature and generally considered a proxy for formation epoch  \citep[although see][]{2018MNRAS.475.4411S,2019MNRAS.485.1906R,2020MNRAS.498.4450W}.

\subsubsection{Halo Spin}
A dimensionless measure of the angular momentum of the halo is 
\begin{equation}
 \lambda = \dfrac{J |E|^{1/2}}{GM_{\rm vir}^{5/2}},
\end{equation}
where $J$ is the magnitude of the angular momentum, $E$ is the total energy, $M_{\rm vir}$ is the virial mass of the halo and $G$ is Newton's constant \citep{1969ApJ...155..393P}.

\subsubsection{Halo Shape}
Dark matter haloes in general have a triaxial shape that can be quantified using the mass ellipsoid tensor.
\begin{equation}
 M_{ij} = \sum_{n \in {\rm halo}}\dfrac{x_{n,i}x_{n,j}}{r_{n}^2} ,
\end{equation}
 where  $\mathbf{x_{n}}$ is the comoving position of the $n^{th}$ particle with respect to the halo center-of-mass and $r_n$ is the corresponding ellipsoidal distance. 
 We can arrange the eigenvalues of $M_{ij}$ as $a^2\geq b^2\geq c^2$ and  characterise the halo shape by the ratio of the smallest to the largest eigenvalue $c/a$. The mass tensor is computed iteratively, each time only including the particles inside the ellipsoid found in the previous step with semi-major axis equal to the virial radius of the halo \citep{2006MNRAS.367.1781A}.

\subsubsection{Velocity ellipsoid}
The velocity ellipsoid for a halo with N particles is given by
\begin{equation}
 V_{ij}^{2}=\sum_{n \in {\rm halo}}\dfrac{v_{n,i}v_{n,j}}{N},
\end{equation}
where $\mathbf{v_{n}}$ is the relative velocity of the $n^{th}$ particle with respect to the bulk velocity of the halo. We can arrange the eigenvalues of $V^{2}_{ij}$ as $a_{v}^2\geq b_{v}^2\geq c_{v}^2$ and characterise the velocity ellipsoid asphericity by the ratio of the smallest to the largest eigenvalue $c_{v}/a_{v}$. This is one measure of the anisotropy in the velocity dispersion of a halo.                      
\subsubsection{Velocity anisotropy}
Another measure of the kinematics of a halo's dark matter content is given by the velocity anisotropy \citep{1987gady.book.....B}.
\begin{equation}
 \beta = 1 - \dfrac{\sigma_{t}^2}{2\sigma_{r}^2}\,,
 \label{eq:beta-def}
\end{equation}
where $\sigma_{t}^2$ and $\sigma_{r}^2$ are the tangential and radial velocity dispersion of the particles in a halo.

We modified \textsc{rockstar} to compute and output the velocity anisotropy $\beta$ and velocity ellipsoid asphericity $c_{v}/a_{v}$. See \citet{2019MNRAS.489.2977R} for further details.           

\subsection{Standardised tidal anisotropy}
\label{subsec:alpha}
We characterise the local environment around the halo by the tidal anisotropy $\alpha$ obtained from the eigenvalues of the tidal tensor Gaussian-smoothed at scale $R_{\rm G} = 4R_{\rm 200b}/\sqrt{5}$ of the halo \citep{2018MNRAS.476.3631P}. The procedure in detail is as follows.

We first compute the overdensity field using the cloud-in-cell (CIC) algorithm on a $1024^3$ grid. Then, a Gaussian kernel is used to smooth the field on various smoothing scales $R_{\rm G}$ starting from the value appropriate for the minimum $R_{\rm 200b}$ in the simulation to $6.5\Mpch$ in 40 logarithmically spaced intervals, which in Fourier space is $\delta(\mathbf{k};R_{\rm G}) = \delta(\mathbf{k})e^{-\mathbf{k}^2R_{\rm G}^2/2}$. The tidal tensor field can now be obtained for a range of smoothing scales $R_{\rm G}$ by inverting the Poisson equation and taking derivatives i.e, the inverse Fourier transform of $(k_{i}k_{j}/k^2)\delta(\mathbf{k};R_{\rm G})$. We evaluate the tidal tensor field at the nearest grid point to the location of halo, linearly interpolated between the two smoothing scales $R_{\rm G}$ closest to $4R_{\rm 200b}/\sqrt{5}$ of the halo; the eigenvalues of this tensor $\lambda_{1},\lambda_{2},\lambda_{3}$ define the halo-centric tidal anisotropy $\alpha$ as follows,
\begin{equation}
\alpha \equiv \sqrt{q^2}/(1+\delta)
\end{equation}
where $q^2=(1/2)[(\lambda_{1}-\lambda_{2})^2+(\lambda_2 - \lambda_3)^2+(\lambda_3 - \lambda_1)^2]$ is the tidal shear and $\delta = \lambda_1 + \lambda_2 +\lambda_3$ is the matter overdensity.

In this work, we rescale (standardise) the tidal anisotropy parameter as follows:
\begin{equation}
\tilde{\alpha} = \dfrac{\ln \alpha - \avg{\ln \alpha |m}}{\sqrt{{\rm Var}(\ln \alpha|m)}},
\label{eq:standardalpha}
\end{equation}
where $\avg{\ln \alpha |m}$  and $\sqrt{{\rm Var}(\ln \alpha|m)}$ are the mean and the central 68.3 percentile of $\ln\, \alpha$ in a narrow bin of mass $m$. In this form, the distribution of $\tilde{\alpha}$ is very well approximated by a standard Gaussian for all $m$ considered (see Figure 1 of \citealp{2020MNRAS.499.4418R}). In practice, we standardise on-the-fly by first calculating the mean and variance in narrow mass bins and then interpolating for the specific halo.

\section{Halo properties from local environment}
\label{sec:propsfromenv}
In this section, we develop an argument for assigning internal halo properties $c$ such as concentration, shape, etc. (see section~\ref{subsec:haloprops}) by knowing a halo's tidal environment $\tilde{\alpha}$ and sampling from an appropriate conditional probability distribution $p(c|m,\tilde{\alpha})$. We first motivate the argument using shuffled measurements of assembly bias and then provide convenient fitting functions for $p(c|m,\tilde{\alpha})$ calibrated using our high- and medium-resolution simulations.

\begin{figure*}
\includegraphics[width=0.9\textwidth]{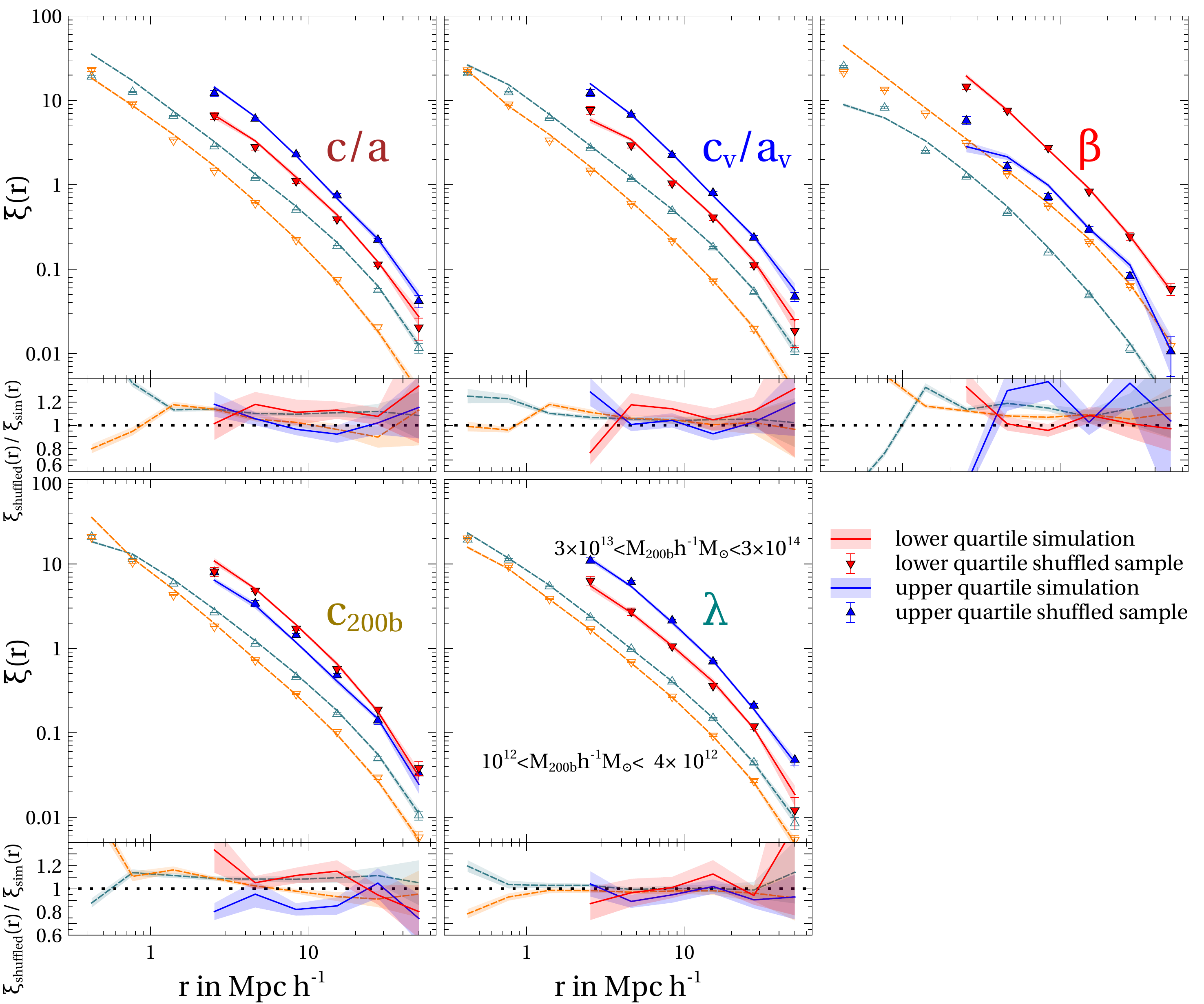}
\caption{Comparison of the real-space 2-point correlation function $\xi(r)$ of haloes in two narrow mass bins measured in simulations (lines) with that obtained from the same haloes shuffled in bins of tidal anisotropy (markers) as described in section~\ref{sec:propsfromenv}. Each panel shows the halo clustering in quartiles of a single halo property (indicated in the label). The solid (dashed) lines show the mass range corresponding to $3\times10^{13}-3\times10^{14}\,(10^{12}-4\times10^{12})\,\Mh$. For halo shape $c/a$, spin $\lambda$ and velocity ellipsoid asphericity $c_v/a_v$, the upper quartile is more clustered than the lower quartile. For velocity anisotropy $\beta$, the lower quartile is more clustered than the upper quartile while the halo concentration $c_{\rm 200b}$ shows opposite trends for the two chosen mass ranges. For each quartile of each property, the shuffled sample accurately reproduces  $\xi(r)$ at separations $1\lesssim r/\Mpch\lesssim50$, as seen in the lower panels which plot the ratio of $\xi(r)$ in the shuffled samples to the corresponding simulation results. All measurements show the average over 10 realisations of our medium-resolution simulation, with error bars and bands showing the standard error in the mean.}
\label{fig:2ptcorr}
\end{figure*}

\subsection{Motivation using assembly bias}
\label{subsec:2ptcorr}
\citet{2019MNRAS.489.2977R} showed that the distribution of any small-scale halo property $c$ (e.g. all those in the previous section), conditioned on the values of large-scale bias $b_1$, the intermediate-scale $\tilde\alpha$, and halo mass $m$, satisfies 
\begin{align}
 p(c|b_1,\tilde{\alpha},m) &= p(c|\tilde{\alpha},m)  \,.
 \label{eq:pcam}
\end{align}
Therefore, the joint distribution of $c$ and $b_1$, conditioned on $\tilde{\alpha}$ and $m$, satisfies 
\begin{align}
  p(c,b_1|\tilde{\alpha},m) &\equiv \frac{p(c,b_1,\tilde{\alpha},m)}{p(\tilde{\alpha},m)}
  = \frac{p(c|b_1,\tilde{\alpha},m)\,p(b_1,\tilde{\alpha},m)}{p(\tilde{\alpha},m)} \nonumber\\
  &= p(c|\tilde{\alpha},m)\,p(b_1|\tilde{\alpha},m),
  \label{eq:indep}
\end{align}
where the first term of the final expression uses equation~(\ref{eq:pcam}).
Assembly bias is a statement about how the correlation between $c$ and $b_1$ depends on $\tilde{\alpha}$ and $m$:
$\avg{c\,b_1|\tilde{\alpha},m}$.   Therefore, if a simulation has already produced the correct spatial distribution and local tidal environment of haloes -- in other words, the distribution $p(b_1,\tilde{\alpha}|m) = p(b_1|\tilde{\alpha},m)\,p(\tilde{\alpha}|m)$ -- then assigning small-scale halo property $c$ using $p(c|\tilde{\alpha},m)$ is \emph{guaranteed} to produce the correct assembly bias for that property.  

The useful consequence of equation~(\ref{eq:indep}), which we exploit below, is this:  Since $\tilde{\alpha}$ is defined at scales larger than the halo radius (section~\ref{subsec:alpha}), it can be accurately estimated for relatively poorly resolved haloes as compared to, say, halo concentration, shape or spin (Appendix~\ref{app:resoln}). This means a sampling of halo property $c$ conditioned on $\tilde{\alpha}$ has the potential to substantially increase the dynamic range of haloes that can be used in a large-volume simulation box. At the low-mass end of typical Gpc-sized simulations \citep{2018ApJS..236...43G,2020ApJS..250....2V}, a factor 10 improvement in mass resolution (which is not unreasonable, as we will demonstrate below) can translate to an increase by factors of $\sim8$-$30$ in halo number, depending on the simulation's particle mass resolution. 

\subsection{A shuffling test of assembly bias}
\label{sec:shuffle}
In this  section, we present a proof-of-concept exercise to establish this idea, using pre-existing measurements of all halo properties computed using \textsc{rockstar}. 
In this exercise,  we sample $p(c|\tilde{\alpha},m)$ by directly shuffling measurements in our medium-resolution simulations to produce `mock' halo properties. Thus, each halo is now endowed with not only its actual value of $c$ as determined by \textsc{rockstar}, but also with a mock value of $c$. However, for the mock value, equation~(\ref{eq:indep}) is true by construction.  Therefore, a comparison of the clustering of haloes selected by their actual and mock $c$ values provides a test of our idea.

To implement the sampling of $p(c|\tilde{\alpha},m)$, we  subdivide haloes into narrow bins in mass $m$ and tidal anisotropy $\tilde{\alpha}$ (we use quintiles of $\tilde{\alpha}$ for each mass bin). Each halo property $c$ is then shuffled amongst the haloes within the same 2-dimensional bin of $(m,\tilde{\alpha})$, thus preserving the correlations between $c$ and $(m,\tilde{\alpha})$. 

Figure~\ref{fig:2ptcorr} shows the real-space 2-point correlation function $\xi(r)$ for sub-populations of haloes in two narrow mass bins as indicated,\footnote{The lower mass range in Figure~\ref{fig:2ptcorr} contains haloes with at least 500 particles, which is sufficient to resolve all five halo properties and their correlations with the local tidal environment (Appendix~\ref{app:resoln}).} and further split into the upper and lower quartiles of each internal halo property (different panels) as generated directly by \textsc{rockstar} (lines) or by shuffling (symbols). We use the natural estimator \citep{1974ApJS...28...19P} of the correlation function,
\begin{equation}
 \xi(r) = DD(r)/RR(r)-1 ,
\end{equation}
which is appropriate for our periodic simulation boxes, where $DD$ is the number of halo pairs with separation in the range $(r,r+\Delta r)$, and $RR$ is the same quantity for a random distribution of the same number density. For $N_{\rm D}$ haloes in a periodic box of side $L_{\rm box}$, $RR = N_{\rm D} (N_{\rm D}/L_{\rm box}^3) \,4\pi r^2\, \Delta r$.

Notice that $\xi(r)$ in the shuffled sample is in excellent agreement 
with the original one, not just on very large scales, but all the way down to the scale on which $\alpha$ is defined, which is approximately $1\Mpch$ for the lower mass range we display.\footnote{Note that, if the shuffling were performed only at fixed mass, we would effectively be sampling the distribution $p(c|m)$, and the resulting mock catalog would \emph{not} recover the assembly bias due to $c$. Instead, halo samples with high and low values of $c$ at fixed mass would have identical clustering strengths (within measurement errors). We have explicitly checked that this does, in fact, happen. This is similar to the `assembly bias erased' mocks discussed by \citet{2007MNRAS.374.1303C,2015MNRAS.451L..45H,2018ApJ...853...84Z}.} 
At separations $\lesssim 1\Mpch$, on the other hand, the shuffled samples differ slightly from the original, indicating that factors other than the tidal environment play a significant role in the spatial correlations between the internal properties of nearby haloes. 

Overall, we conclude that equation~(\ref{eq:indep}), and hence our basic concept of sampling distributions of halo properties conditioned only on halo mass and local environment, is remarkably successful at reproducing clustering, within 10\% accuracy at most scales and within 20\% in the worst case scenario on scales $\gtrsim\textrm{few}\,\Mpch$ . 

\subsection{Probability distribution of halo properties}

For the results of the previous subsection to be useful in practice, we also need $p(c|m,\tilde{\alpha})$.  In principle, we could simply measure these conditional distributions in our medium- and high-resolution simulations and provide them in tabular form.  However, if $p(c|m,\tilde{\alpha})$ were Gaussian, then we would simply need to describe how the mean and variance depend on $m$ and $\alpha$.  Moreover, recalling that $\tilde{\alpha}$ is approximately a standard Gaussian variate (section~\ref{subsec:alpha}), the description of the conditional distribution $p(c|m,\tilde{\alpha})$ greatly simplifies if $c$ is \emph{also} approximately Gaussian distributed.  In this case, in fixed mass bins (and suppressing the explicit mass dependence for brevity), we would have
\begin{align}
p(c|\tilde{\alpha})&=\dfrac{e^{-(c-\rho_{c} \tilde{\alpha}-\mu_{c} )^2/2\sigma_{c}^2(1-\rho_{c}^2)}}{\sqrt{2 \pi\sigma_{c}^2(1-\rho_{c}^2)}}\,.
\label{eq:probdist}
\end{align}
Here, $\mu_{c}$ and $\sigma_{c}$ are the mean and the standard deviation of the marginal distribution $p(c)$ and $\rho_{c}$ is the correlation coefficient between $c$ and $\tilde{\alpha}$.

\begin{figure}
\includegraphics[width=0.95\linewidth]{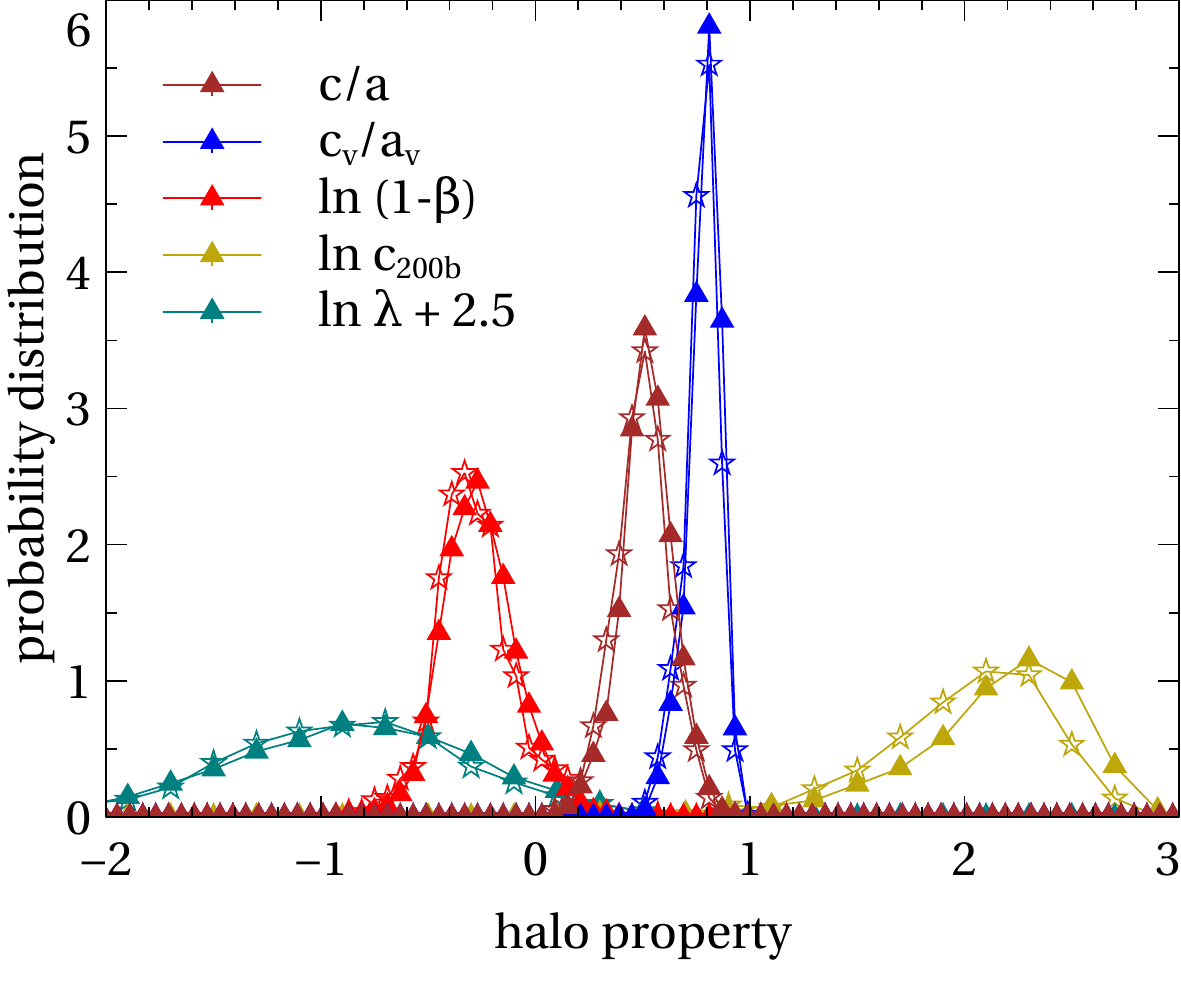} 
\caption{Probability distribution of all the five halo properties, transformed in each case to be close to Gaussian distributed (halo shape $c/a$ and velocity ellipsoid asphericity $c_v/a_v$ are not transformed; see text for a discussion). Triangle (star) markers show the distribution for the mass range 1-2 (4-6)$\times 10^{13}\Mh$. Similar results hold for all the mass bins we consider in this work. For halo spin $\lambda$, we added an arbitrarily chosen constant 2.5 to $\ln\lambda$ for visual clarity.}
\label{fig:probdist}
\end{figure}    

After some exploration (Appendix~\ref{app:haloprops}), we decided to approximate the distribution of the halo properties as follows,
\begin{itemize}
 \item halo concentration $c_{\rm 200b}\rightarrow$ Lognormal 
 \item halo spin $\lambda \rightarrow$ 
 Lognormal
 \item halo shape $c/a \rightarrow$ Gaussian 
 \item halo velocity ellipsoid $c_{v}/a_{v}\rightarrow$ Gaussian
 \item velocity anisotropy $\beta \rightarrow$ Gaussian in $\ln(1-\beta)$
\end{itemize}
Figure~\ref{fig:probdist} shows the measured probability distributions of all the halo properties that we approximate as Gaussian, for a couple of narrow mass bins.

\begin{figure*}
 \includegraphics[width=0.95\linewidth]{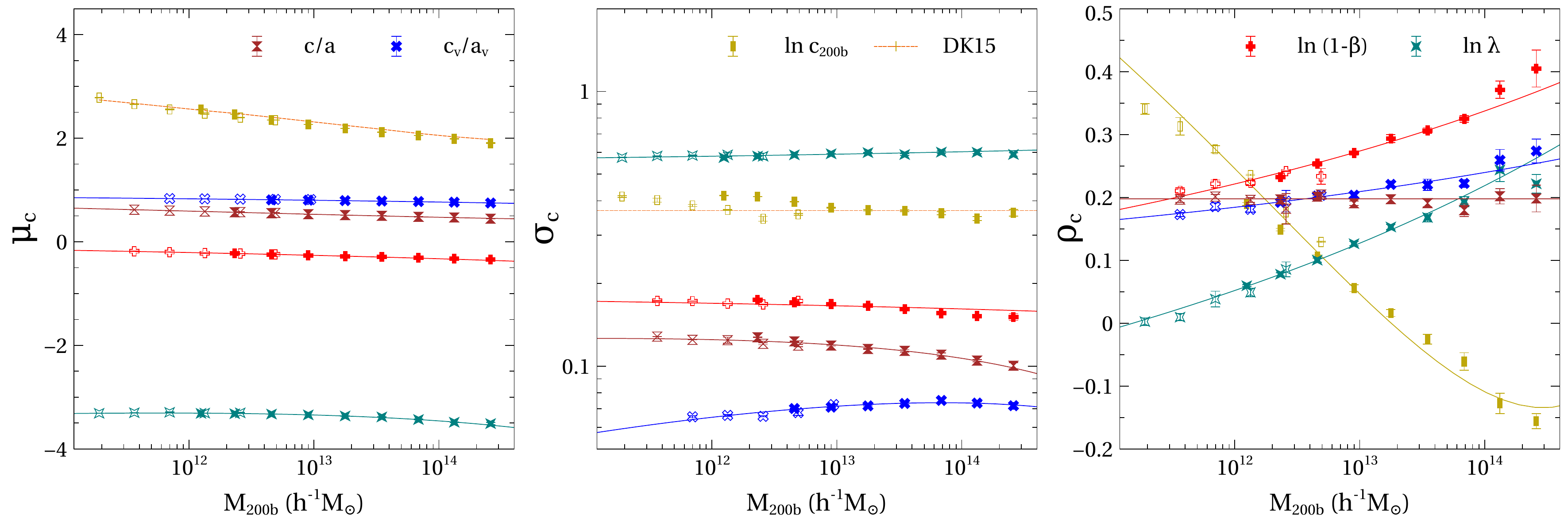}
 \caption{Summary statistics of the halo properties as a function of halo mass: median halo property $\mu_{c}$ \emph{(left panel)},  central 68.3 percentile of the halo property $\sigma_{c}$ \emph{(middle panel)} and the correlation coefficient $\rho_{c}$ of the halo property with standardised tidal anisotropy $\tilde{\alpha}$ \emph{(right panel)}. The empty (filled) markers are from the high-resolution (medium-resolution) simulations. Solid lines show polynomial fits to the data in the variable $\ln\nu$ (Tables~\ref{table:ca}-\ref{table:c200b}), except for $\mu_{\ln c_{\rm 200b}}$ and $\sigma_{\ln c_{\rm 200b}}$ where we show the calibrations from \citet[][DK15]{2015ApJ...799..108D}. These curves contain all the information required to generate realistic halo properties endowed with accurate halo assembly bias, by sampling the probability distribution function $p(c|m,\tilde{\alpha})$ from \eqn{eq:probdist} which can be constructed knowing $\mu_c$, $\sigma_c$ and $\rho_c$ for each property $c$. For reference, the left and right edges of each plot, namely the mass values $1.2\times10^{11}\Mh$ and $4\times10^{14}\Mh$, respectively correspond to $\ln\nu$ values of $-0.47$ and $0.94$ for our cosmology.}
 \label{fig:fits}
\end{figure*}

In Figure~\ref{fig:fits}, the symbols in the \emph{left}, \emph{middle} and \emph{right panels} show measurements from our medium- and high-resolution simulations of $\mu_{c},\sigma_{c}$ and $\rho_{c}$, respectively, as a function of halo mass for each of the (Gaussianised) halo properties $c$.  For the properties $c/a$ and $c_v/a_v$ which are already treated as Gaussian distributed, we estimate $\rho_c$ using Pearson's correlation coefficient. However, as discussed by \citet{2020MNRAS.499.4418R}, we note that Pearson's coefficient is sensitive to outliers/tails in the distribution and cannot be reliably used for those halo properties with skewed distributions, namely halo spin $\lambda$, concentration $c_{\rm 200b}$ and velocity anisotropy $\beta$. For these properties, we therefore follow the prescription outlined by  \citet{2020MNRAS.499.4418R} and first compute Pearson's correlation for the variables $\lambda$, $c_{\rm 200b}$ and $1-\beta$, which are approximately Lognormal distributed, and then convert it to the correlation between their Gaussian counterparts using the analytical relation between Lognormal and Gaussian correlation coefficients \citep[equation~B2 of][]{2020MNRAS.499.4418R}.
The mass ranges for the measurements in each simulation volume were chosen based on the resolution study presented in Appendix~\ref{app:resoln}. 

The solid curves  for $\mu_c$ and $\sigma_c$ for $c=\ln c_{\rm 200b}$ are obtained from the fits for the median and variance of $\ln c_{\rm 200c}$ provided by \cite{2015ApJ...799..108D}, converted to the $200{\rm b}$ definition assuming NFW profiles using the prescription of \citet{2003ApJ...584..702H}. These accurately describe the corresponding measurements in our simulations.
Similar simple fitting functions were not available in the literature for the other variables, and we have calibrated them ourselves.
The solid curves for the remaining variables are polynomial fits (Tables~\ref{table:ca}-\ref{table:c200b}) in the variable $\ln(\nu)$.\footnote{Here $\nu=1.686/\sigma(m)$ and $\sigma^2(m)$ is the variance of linear density fluctuations, extrapolated to $z=0$ and smoothed with a spherical tophat kernel with Lagrangian radius $R = (3m/4\pi G\bar\rho)^{1/3}$, with $\bar\rho$ being the mean density of the Universe at $z=0$.} The degree of the polynomial to be used for each variable was determined by an analysis with the Akaike Information Criterion with correction (AICC).\footnote{
The Akaike information criterion with correction (\citealp{1974ITAC...19..716A,sugiura}; see \citealp{2007MNRAS.377L..74L} for a review) is given by ${\rm AICC}=\chi^2_{\rm min}+2kN/(N-k-1)$, where $\chi^2_{\rm min}$ is the minimum Chi-squared of the fit, $N$ is the number of data points and $k$ is the number of fitted parameters. Due to small error bars in our data, the minimum Chi-squared values are typically large which causes the AICC to decrease with increasing degree of polynomial, with no clear minimum in many cases. To robustly select the best description of each data set, we progressively increase the degree of the polynomial being tested until two criteria are satisfied: (a) the residuals of the polynomial compared to the data are smaller than $\sim10\%$ over the range of the data and (b) the AIC decreases substantially as compared to the previous (lower degree) polynomial.}

Given a halo catalog at $z=0$ with measured values of $M_{\rm 200b}$ and $\alpha$ for each halo, internal properties $c$ can then be assigned one at a time by (i) standardising $\alpha$ in narrow mass bins to obtain $\tilde{\alpha}$, and (ii) using the fits for $\mu_{c},\sigma_{c}$ and $\rho_{c}$ shown in Figure~\ref{fig:fits} and Tables~\ref{table:ca}-\ref{table:c200b} to sample the conditional (1-dimensional) distributions $p(c|\tilde{\alpha})$ (equation~\ref{eq:probdist}) and assign values of $c$ to individual haloes in each mass bin. We note that the fits collectively span between 3 to 4 orders of magnitude in mass (with some variation depending on halo property). 
In the next section, we apply this technique to our low-resolution simulation to demonstrate its power in extending the dynamic range of the simulation boxes.

\section{Application to large-volume simulations}
\label{sec:application}

As noted in the Introduction, resolving low-mass objects in large volumes is difficult. The internal properties of objects resolved with $\lesssim\textrm{few}\times100$ particles are known to be affected by numerical resolution effects. Since the tidal anisotropy is computed at a larger scale ($\sim\textrm{few}\times$ halo radius) than the internal halo properties, one might expect it to have better convergence behaviour than the internal properties at the same mass. If this holds, then the technique outlined above is potentially a powerful method to increase the available dynamic range in halo properties in a large-volume, low resolution simulation. We demonstrate below that this is indeed the case.

\begin{figure*}
\includegraphics[width=0.8\linewidth,trim=0 10 5 5,clip]{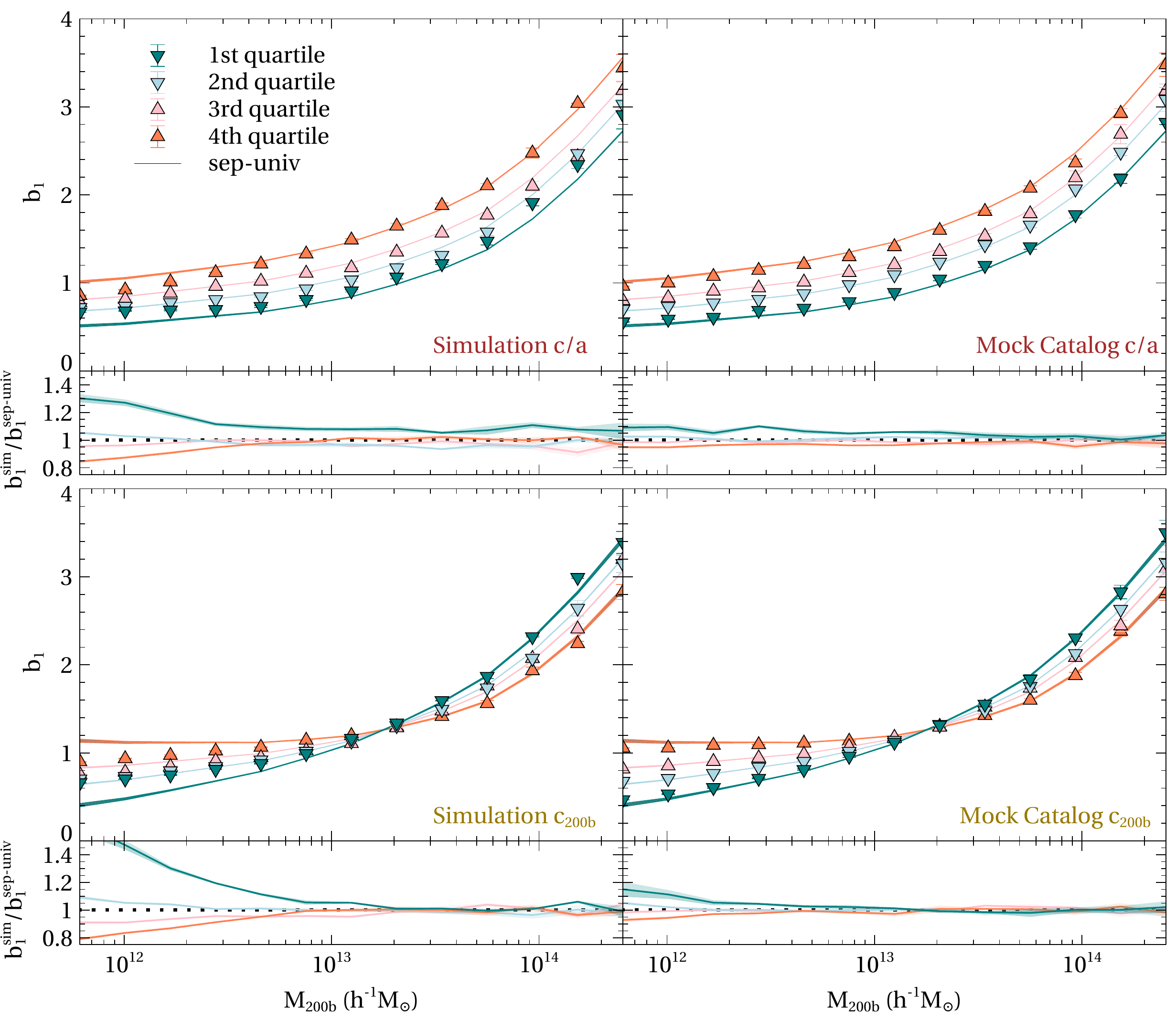} 
\caption{Comparison of assembly bias due to halo shape $c/a$ \emph{(top row)} and concentration $c_{\rm 200b}$ \emph{(bottom row)} in our large-volume, low-resolution simulations and corresponding mocks with Separate Universe (SU) calibrations. For each halo property $c$, the four different coloured data markers show the average halo bias in the four quartiles of $c$, with values $c$ measured directly in the low-resolution simulation \emph{(left panels)} and assigned by our conditional sampling algorithm \emph{(right panels)}. The solid curves, repeated for each property in the corresponding left and right panel, show the calibration for assembly bias using the SU technique from equation 27 of \citet{2020MNRAS.499.4418R}. The small lower sub-panels in each case show the ratio of each assembly bias curve with the SU calibration. Low-mass haloes resolved with $\lesssim300$ particles (halo masses $\lesssim 4.6\times 10^{12}h^{-1}M_\odot$) in the simulation (left hand panels) fail to reproduce the full strength of assembly bias, while corresponding haloes in the mock catalogs (right hand panels) perform much better down to a 30 particle threshold. 
}
\label{fig:largevolmocksu1}
\end{figure*}
 
 \begin{figure*}
\includegraphics[width=0.8\linewidth,trim=0 10 5 5,clip]{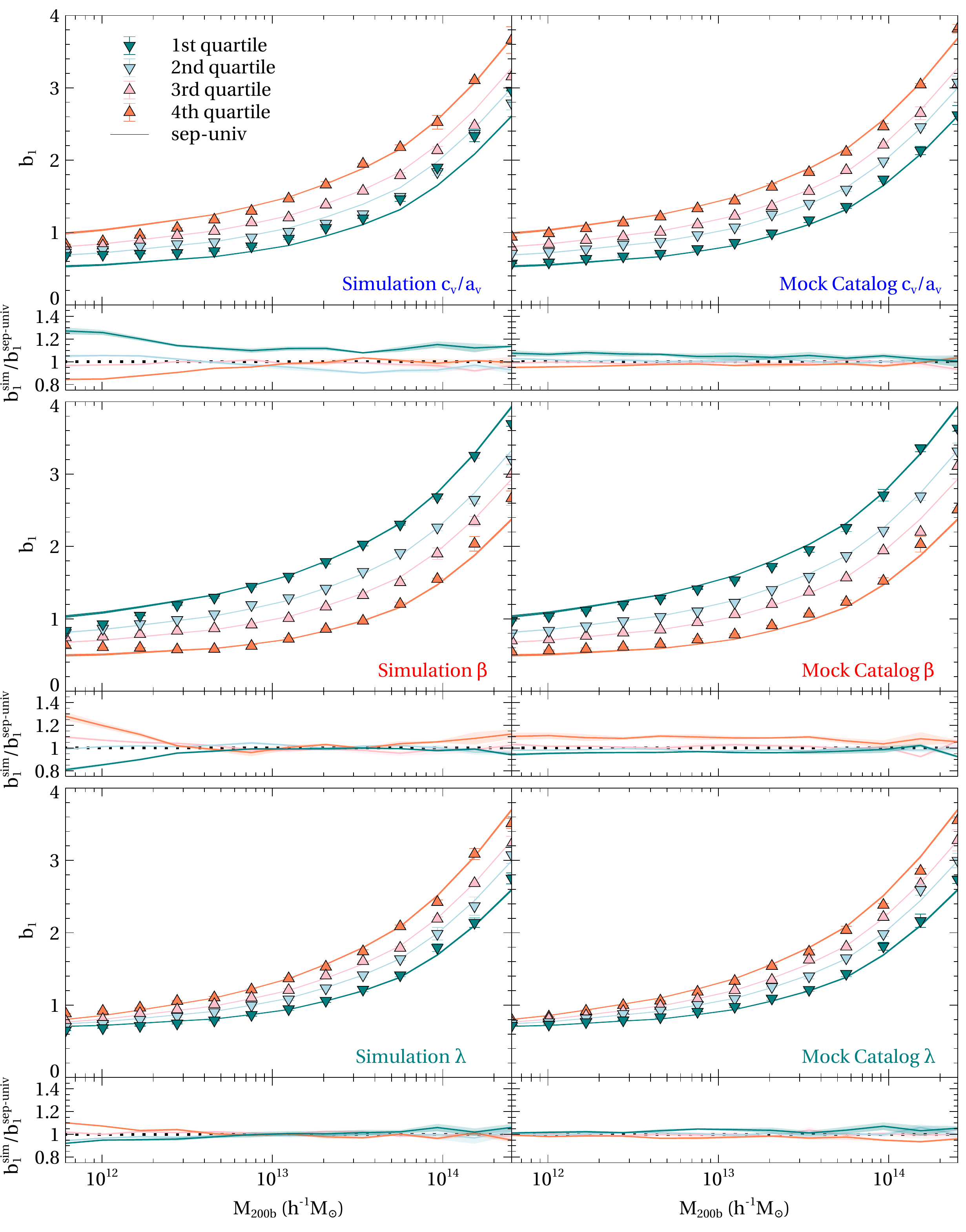} 
\caption{Same as Figure~\ref{fig:largevolmocksu1}, showing results for velocity ellipsoid asphericity $c_{v}/a_{v}$ \emph{(top row)}, velocity anisotropy $\beta$ \emph{(middle row)} and halo spin $\lambda$ \emph{(bottom row)}. Similarly to halo shape and concentration, we again see that our mock algorithm outperforms the actual simulation in reproducing assembly bias trends for poorly resolved haloes.}
\label{fig:largevolmocksu2}
\end{figure*}

In Appendix~\ref{app:resoln}, we present a resolution study using our medium- and high-resolution boxes to determine the minimum particle count required for convergence of the basic variables $\mu_c$, $\sigma_c$ and $\rho_c$ for different halo properties $c$ used in this work, compared with a similar exercise for the median and variance of $\ln\alpha$. The bottom-line of this study is that the required statistics of the halo properties are typically well-resolved only for haloes with particle counts $\gtrsim500$ \citep[consistent with previous studies, see e.g.][]{2007MNRAS.376..215B,2017MNRAS.468.2984P,2020MNRAS.500.3309M}, while the median and variance of $\alpha$ are accurately resolved for haloes containing as few as 30 particles. Thus, our methodology can lead to potential gains of more than an order of magnitude in mass.

To demonstrate this explicitly, we generate mock values for haloes in our large-volume (600$\Mpch$), low-resolution simulation box (in which 300~particles corresponds to a mass of $4.6\times 10^{12}\Mh$) by sampling from 
$p(c|m,\tilde{\alpha})$ and measuring the resulting clustering signal.  
Figures~\ref{fig:largevolmocksu1} and~\ref{fig:largevolmocksu2} compare the corresponding assembly bias with that measured in the original (resolution compromised) simulation for a range of choices for $c$.  (See Appendix~\ref{sec:measuringbias} for details of the measurement procedure.) As  reference values, we use the calibrations of large-scale, scale-independent assembly bias from the Separate Universe (SU) technique presented in \citet{2020MNRAS.499.4418R} which take as input the correlation between Gaussianized $c$ and $\tilde{\alpha}$. Besides being a different approach to measuring halo bias, the SU calibrations are both very accurate and not limited by box-size effects. One may also treat direct $b_{1}$ measurements from a higher resolution simulation as the reference. This comparison is shown in Figure~\ref{fig:largevolmock}:  the mock values are closer to the high-resolution curves.  However, since the high-resolution simulation is a smaller volume box, the bias measurements suffer from systematic effects due to missing long-wavelength modes which affect the magnitude of $b_1$ for individual haloes. Due to this systematic ambiguity, and because the SU calibration has very low noise, we prefer to use the SU results as our reference.

The panels on the left show that, typically, the results in the original low-resolution simulation show a smaller assembly bias signal than the SU reference. This is easily understood by inspecting Figure~\ref{fig:convergencestudy2}, which shows that resolution artefacts tend to degrade the correlations between halo properties and $\alpha$, whereas the statistics of $\alpha$ itself are accurately estimated at the same masses.  In contrast, our mock sampling technique does not suffer from this drawback, since it is built using fitting functions that used well-resolved haloes in high-resolution simulations to access low halo masses.  

The panels on the right show that our mocks are clearly closer to the reference (SU) than are the original simulations, especially at low masses.  The mocks typically agree with the SU calibration to better than $\sim5\%\,(\sim15\%)$ for the middle (outer) quartiles of \emph{all} halo properties down to the 30 particle limit of the simulation.  For halo  spin $\lambda$, the mocks are within a few per cent of the SU reference for \emph{all} quartiles at all masses.  Additionally, for haloes resolved with $\lesssim300$ particles, the mocks substantially outperform the raw (resolution compromised) measurements.  Together, we believe that Figures~\ref{fig:largevolmocksu1} and~\ref{fig:largevolmocksu2} demonstrate the power and accuracy of our approach.

\section{Summary}
\label{sec:summary}
We have explored the idea that the knowledge of the mass $m$ and local tidal environment (characterised by the standardised local tidal anisotropy $\tilde{\alpha}$, see equation~\ref{eq:standardalpha}) of a population of haloes in an $N$-body simulation is sufficient to produce accurate statistical estimates of several \emph{internal} halo properties $c$ such as concentration, shape, spin and variables related to velocity dispersion structure, by sampling appropriate conditional distribution functions $p(c|m,\tilde{\alpha})$. 

This not only leads to the correct overall distribution of each property $c$ for the full halo population, but also correctly reproduces the dependence of \emph{large-scale} clustering on $c$, namely assembly bias. The crux of our idea is the recognition that $\tilde{\alpha}$ statistically explains the assembly bias signal of a number of halo properties \citep{2019MNRAS.489.2977R}.
Our main results are as follows.
\begin{itemize}
  \item As a proof-of-concept, we used well-resolved haloes at $z=0$ in a medium-resolution simulation to show that shuffling halo properties $c$ in narrow bins of mass and tidal anisotropy accurately preserves assembly bias in the real-space 2-point clustering of haloes all the way down to separations of $\sim1\Mpch$ (section~\ref{subsec:2ptcorr}, Figure~\ref{fig:2ptcorr}). 
  \item Using medium- and high-resolution simulations, we calibrated the conditional distributions $p(c|m,\tilde{\alpha})$ (equation~\ref{eq:probdist}) of five (Gaussianised) halo properties $c$, using convenient fitting functions  over a wide range of halo masses at $z=0$ (Tables~\ref{table:ca}-\ref{table:c200b}, Figure~\ref{fig:fits}). Since $\tilde{\alpha}$ is a standard Gaussian variate for each mass bin, these fits can also be easily adapted to produce the \emph{unconditional} distributions $p(c|m)$ at fixed halo mass.
  \item We showed that a straightforward sampling of the (1-dimensional) conditional distributions \eqref{eq:probdist} for haloes with measured values of mass and $\tilde{\alpha}$ in a \emph{large-volume, low-resolution} simulation leads to `mock' halo properties $c$ that accurately reproduce large-scale assembly bias for haloes resolved with as few as 30 particles (Figures~\ref{fig:largevolmocksu1} and~\ref{fig:largevolmocksu2}), thus increasing the available dynamic range of the simulation by more than an order of magnitude in halo mass.
 \end{itemize}

It is worth bearing in mind that, although we have focused on accurately modelling statistical correlations between halo properties and the larger-scale halo environment, we have not tried to accurately describe all the properties of an individual halo simultaneously. For example, we did not show how to include correlations between pairs of halo properties.  However, the extension of our logic is conceptually straightforward:  the scalar halo property $c$ must be replaced by a vector $\mathbf{c}$.  Determining the  combinations of the components of $\mathbf{c}$ that provide an optimal description of assembly bias effects is then a very interesting avenue to explore \citep{lms17,han+19}, with natural connections to machine learning applications for generating mock galaxy catalogs \citep{2018MNRAS.478.3410A,2020arXiv201200111W}.  

In addition, we focused on a single cosmology at a single redshift.  The cosmology and redshift dependence of the conditional distributions $p(c|\alpha,m)$ remains to be calibrated. However, this does not pose a fundamental challenge to the conceptual underpinning of our approach, namely, that knowledge of the local (intermediate-scale) environment of haloes can lead to accurate statistical representations of internal (small-scale) halo properties. In future work, we will extend our technique to modelling multiple variables as well as exploring the redshift and cosmology dependence of the calibrations presented here.  We also plan to explore the performance of our technique in reproducing clustering statistics beyond the 2-point function.

Finally, although we have show-cased the power of our technique for improving an $N$-body halo catalog at the low-mass end, the fitting functions we provide in Appendix~\ref{app:haloprops} perform accurately over several orders of magnitude in halo mass (Figures~\ref{fig:largevolmocksu1} and~\ref{fig:largevolmocksu2}). This opens up the exciting new possibility of applying our technique in fast approximate simulation methods \citep{2013MNRAS.433.2389M,2013MNRAS.435L..78K,2013JCAP...06..036T,2014MNRAS.437.2594W,2015MNRAS.450.1856A,2015A&C....12..109H}, which routinely supplant full $N$-body simulations for generating large-volume mock catalogs supporting large-scale galaxy surveys. These fast algorithms are typically calibrated to produce mass-selected clustering statistics comparable to that of full $N$-body simulations. 
Since our technique only requires knowledge of halo positions, masses and the dark matter density field, it can augment the haloes from these algorithms with internal properties that correlate correctly with the surrounding large-scale structure.  As a specific example, our methodology can endow the fast weak lensing simulations of \cite{2020MNRAS.496.1307G} with assembly bias functionality at negligible additional computational cost.  We will report on such applications in future work.  

\section*{Acknowledgments}
We thank the Munich Institute for Astro- and Particle
Physics (MIAPP) and the organisers of the programme on
Dynamics of Large-Scale Structure (July 2019) for their hospitality while this work was conceptualised, Idit Zehavi for comments on an earlier draft and the anonymous referee for a beneficial report.
The research
of AP is supported by the Associateship Scheme of ICTP,
Trieste and the Ramanujan Fellowship awarded by the Department of Science and Technology, Government of India. We gratefully acknowledge the use of high performance computing facilities at IUCAA, Pune.\footnote{\url{http://hpc.iucaa.in}}

\section*{Data Availability}
No new data were generated in support of this research. The simulations used in this work are available from the authors upon reasonable request. 

\bibliography{reference.bib}

\appendix
\section{Distribution of Halo Properties}
\label{app:haloprops}

\begin{figure}
\includegraphics[width=0.95\linewidth]{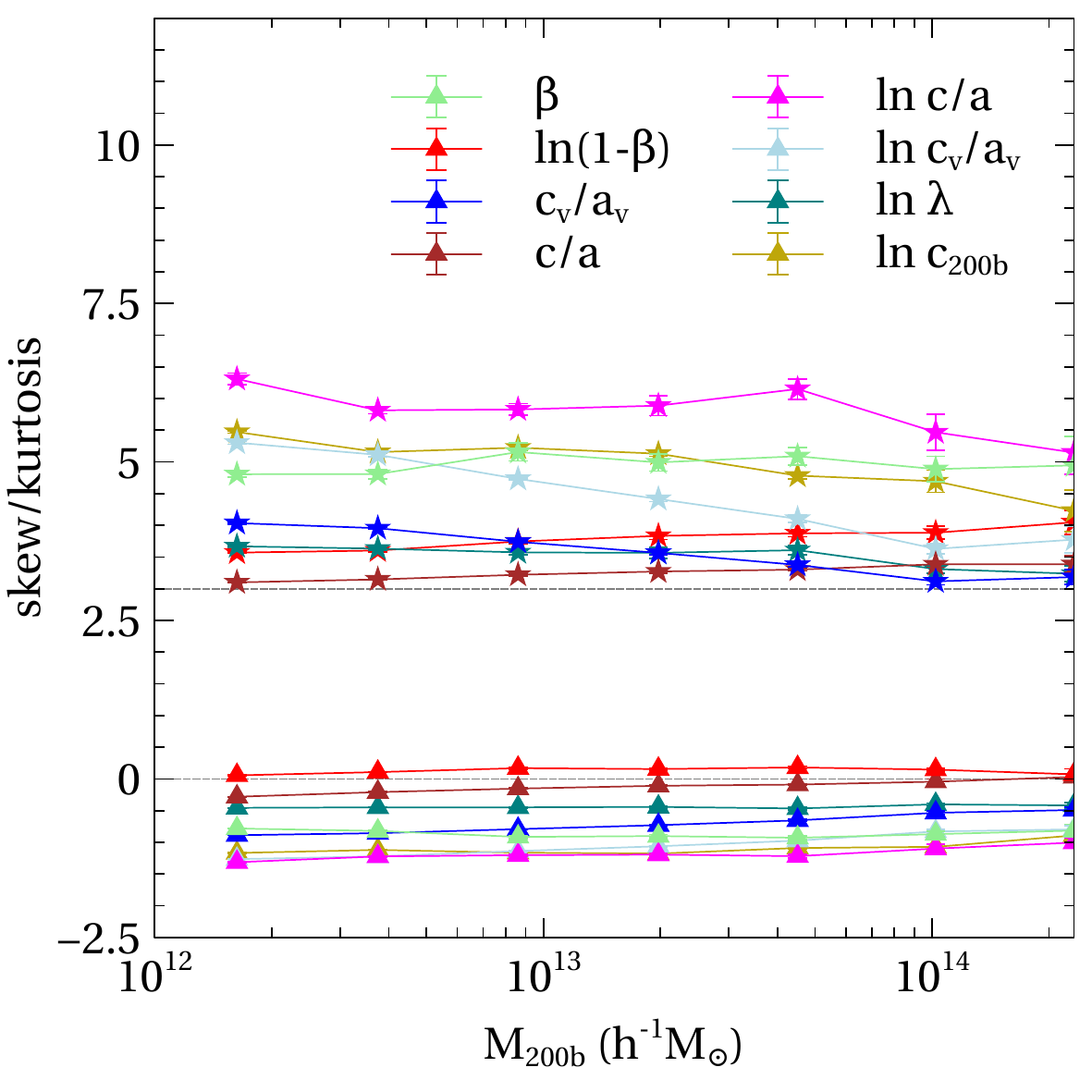} 
\caption{Skewness (triangles) and kurtosis (stars) for the distribution of the halo properties as a function of halo mass. The closeness of the skewness (kurtosis) to the value 0 (3) can be used to assess how closely the corresponding distribution is to a Gaussian. Based on these results, we conclude that $c/a$, $c_v/a_v$, $\ln(1-\beta)$, $\ln\lambda$ and $\ln c_{\rm 200b}$ can be reasonably approximated as being Gaussian distributed (see Appendix~\ref{app:haloprops} for details).}
\label{fig:skewkurt}
\end{figure}

\begin{table}
\centering
 \caption{Best fit coefficients and covariance matrices for the polynomial fits of the mean $\mu_{c}$, variance $\sigma_c$ and Pearson coefficient $\rho_c$ (the correlation with $\tilde{\alpha}$) for the halo property $c=c/a$. In each sub-table, the caption shows the polynomial form as a function of $\ln \nu$, the first row gives the best fit values, the second row gives the standard deviation, and the last few rows give the correlation coefficients.}
  \label{table:ca}
 \subcaption{$\mu_{c/a}={(c/a)}_{0}+{(c/a)}_{1}\ln \nu+{(c/a)}_{2}(\ln \nu)^2$}
 \input{./tables/c_to_a_mean.tex}
 \subcaption{$\sigma_{c/a}=\sigma^{c/a}_{0}+\sigma^{c/a}_{1}\ln \nu+\sigma^{c/a}_{2}(\ln \nu)^2$}
 \input{./tables/c_to_a_sig.tex}
  \subcaption{$\rho_{c/a}=\rho^{c/a}_{0}$}
 \input{./tables/c_to_a_corr.tex}
\end{table}

\begin{table}
\centering
 \caption{Best fit coefficients and covariance matrices for $c_{v}/a_{v}$.}
  \label{table:vca}
 \subcaption{$\avg{c_{v}/a_{v}}={(c_{v}/a_{v})}_{0}+{(c_{v}/a_{v})}_{1}\ln \nu$}
 \input{./tables/vc_to_va_mean.tex}
 \subcaption{$\sigma^{c_{v}/a_{v}}=\sigma^{c_{v}/a_{v}}_{0}+\sigma^{c_{v}/a_{v}}_{1}\ln \nu+\sigma^{c_{v}/a_{v}}_{2}(\ln \nu)^2$}
 \input{./tables/vc_to_va_sig.tex}
  \subcaption{$\rho^{c_{v}/a_{v}}=\rho^{c_{v}/a_{v}}_{0}+\rho^{c_{v}/a_{v}}_{1}\ln \nu$}
 \input{./tables/vc_to_va_corr.tex}
\end{table}

\begin{table}
\centering
 \caption{Best fit coefficients and covariance matrices for $\ln (1-\beta)$.}
  \label{table:beta}
 \subcaption{$\mu_{\ln (1-\beta)}={\beta}_{0}+{\beta}_{1}\ln \nu$}
 \input{./tables/beta_mean.tex}
 \subcaption{$\sigma_{\ln (1-\beta)}=\sigma^{\beta}_{0}+\sigma^{\beta}_{1}\ln \nu$}
 \input{./tables/beta_sig.tex}
  \subcaption{$\rho_{\ln (1-\beta)}=\rho^{\beta}_{0}+\rho^{\beta}_{1}\ln \nu$}
 \input{./tables/beta_corr.tex}
\end{table}

\begin{table}
\centering
 \caption{Best fit coefficients and covariance matrices for $\ln \lambda$.}
  \label{table:spin}
 \subcaption{$\mu_{\ln \lambda}={\lambda}_{0}+{\lambda}_{1}\ln \nu+{\lambda}_{2}(\ln \nu)^2$}
 \input{./tables/spin_mean.tex}
 \subcaption{$\sigma_{\ln \lambda}=\sigma^{\lambda}_{0}+\sigma^{\lambda}_{1}\ln \nu$}
 \input{./tables/spin_sig.tex}
  \subcaption{$\rho_{\ln \lambda}=\rho^{\lambda}_{0}+\rho^{\lambda}_{1} \ln \nu$}
 \input{./tables/spin_corr.tex}
\end{table}

\begin{table}
\centering
 \caption{Best fit coefficients and covariance matrix for $\ln c_{200b}$ \\ $\rho_{\ln c_{200b}}=\rho^{c_{200b}}_{0}+\rho^{c_{200b}}_{1} \ln \nu+\rho^{c_{200b}}_{2} (\ln \nu)^2+\rho^{c_{200b}}_{3} (\ln \nu)^3$.}
  \label{table:c200b}
 \input{./tables/c200b_corr.tex}
\end{table}

\noindent
For the analysis in this paper, we require the form of each halo property that has a distribution closest to a Gaussian. Hence, in Figure~\ref{fig:skewkurt}, we compute the skewness and kurtosis for several forms of the halo property and compare it to those expected from a Gaussian, i.e, 0 and 3 respectively. 
Our choices are summarized below.
\begin{itemize}
    \item \emph{Halo spin $\lambda$ and concentration $c_{\rm 200b}$:} These distributions are well-studied and known to be approximated by a Lognormal form \citep{2001ApJ...555..240B,2005ApJ...627..647B,2015ApJ...799..108D}; this is also apparent from Figure~\ref{fig:skewkurt}, although we note that the approximation is decidely worse for $c_{\rm 200b}$. 
    \item \emph{Halo shape $c/a$:} Some previous studies have argued for using a Lognormal distribution for $c/a$ \citep{2015MNRAS.449.3171B,2017MNRAS.467.3226V}. We find, however, that  $c/a$ has a skewness (kurtosis) substantially closer to 0 (3) as compared to $\ln c/a$. We therefore choose to approximate $c/a$ as Gaussian distributed. 
    \item \emph{Velocity ellipsoid asphericity $c_v/a_v$:} This distribution less studied in the literature. Comparing the skewness and kurtosis of $c_{v}/a_{v}$ with that of $\ln c_{v}/a_{v}$, we conclude that although neither is very close to Gaussian, $c_v/a_v$ is somewhat closer to a Gaussian than its logarithm. We therefore choose to model the $c_{v}/a_{v}$ distribution as Gaussian.
    \item \emph{Velocity anisotropy $\beta$:} We find that the distribution of $\beta$ has a substantial tail towards negative values (not shown). Since $\beta<1$ by construction (see equation~\ref{eq:beta-def}) we tested whether this tail can be accounted for using the transformation $\beta\to\ln(1-\beta)$. Indeed, we see in Figure~\ref{fig:skewkurt} that $\ln(1-\beta)$ has skewness (kurtosis) much closer to 0 (3) than does $\beta$ (see also Figure~\ref{fig:probdist}). Consequently, we model $\ln(1-\beta)$ as being Gaussian distributed.
\end{itemize}

\section{Resolution Study}
\label{app:resoln}

\begin{figure}
  \centering
  \includegraphics[width=0.95\columnwidth]{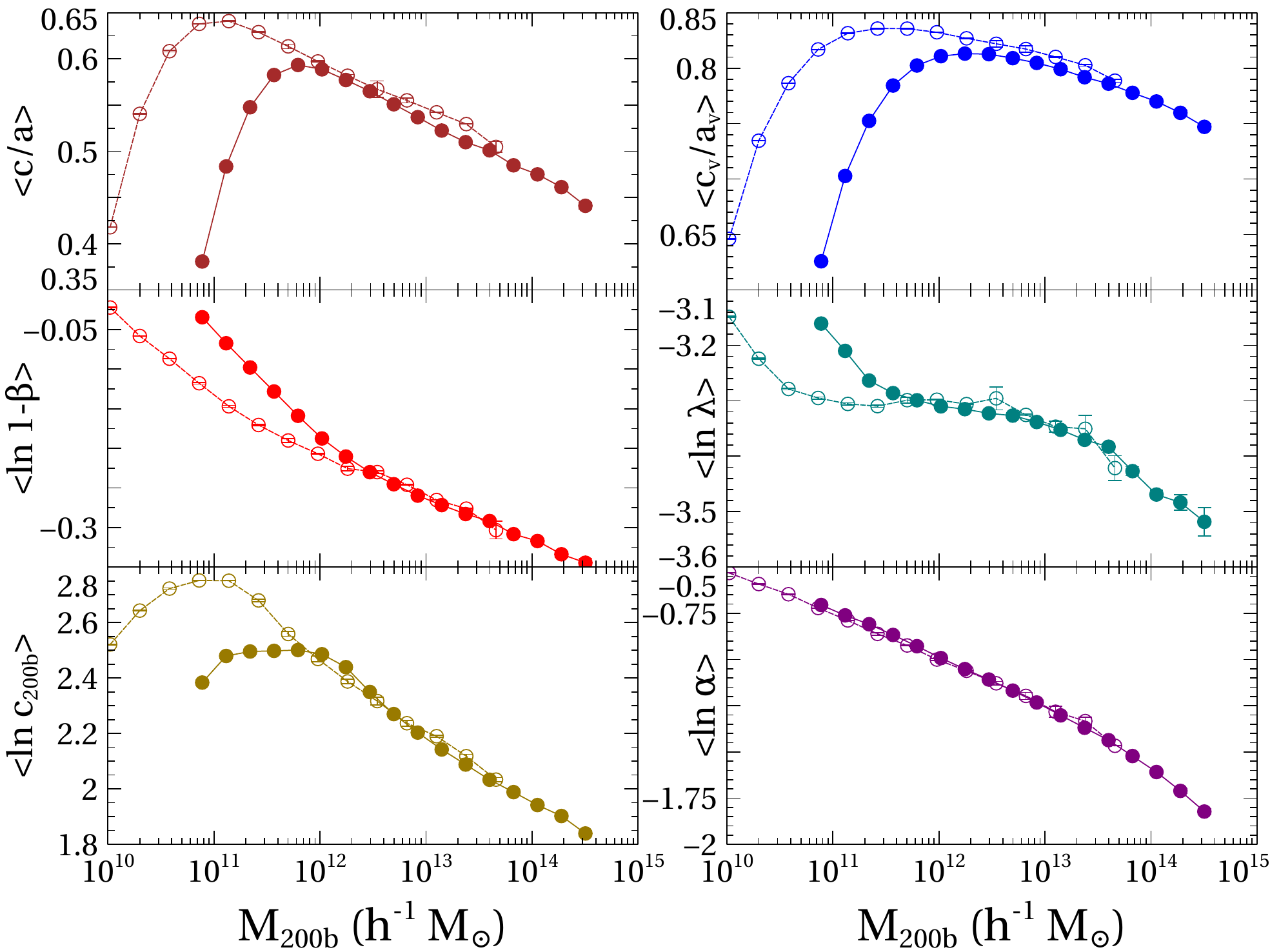}
 \caption{Numerical convergence of the median of the halo properties with particle count. Each panel corresponds to one of the properties we model as being Gaussian distributed (see Appendix~\ref{app:haloprops}), with the bottom right panel showing the results for $\ln\alpha$. The filled (empty) markers correspond to measurements in the medium-resoution (high-resolution) simulation boxes of size 300 (150) $\Mpch$.  The upturn in $\avg{\ln\lambda}$ for $N_{\rm p}\lesssim300$ in each simulation box is extensively studied in \citep{2007MNRAS.376..215B}. The variable $\ln (1-\beta)$ shows a similar upturn at $\sim1000$ particles, while all other variables show a downturn at particle counts between $\sim300$-$1000$. In the bottom right panel, we can see that the median tidal anisotropy around a halo is convergent even for 30 particle haloes.}
  \label{fig:convergencestudy0}
\end{figure}

\begin{figure}
  \centering
  \includegraphics[width=0.95\columnwidth]{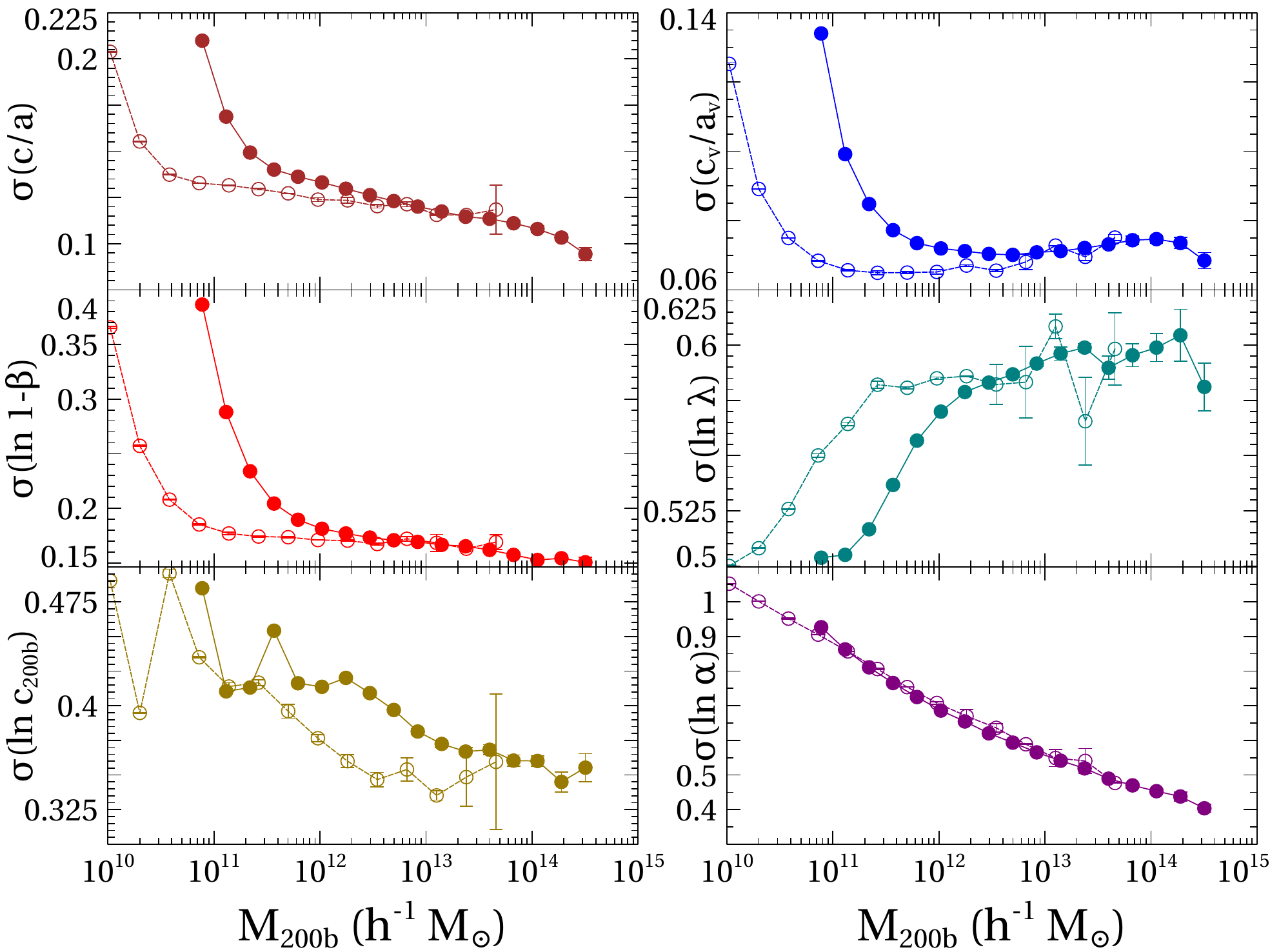}
 \caption{Same as Figure~\ref{fig:convergencestudy0}, showing results for the standard deviation of each halo property and $\ln\alpha$. In the bottom right panel, we can see that the standard devation of tidal anisotropy around a halo is convergent even for 30 particle haloes.}
    \label{fig:convergencestudy1}
\end{figure}

\begin{figure}
  \centering
  \includegraphics[width=0.9\columnwidth]{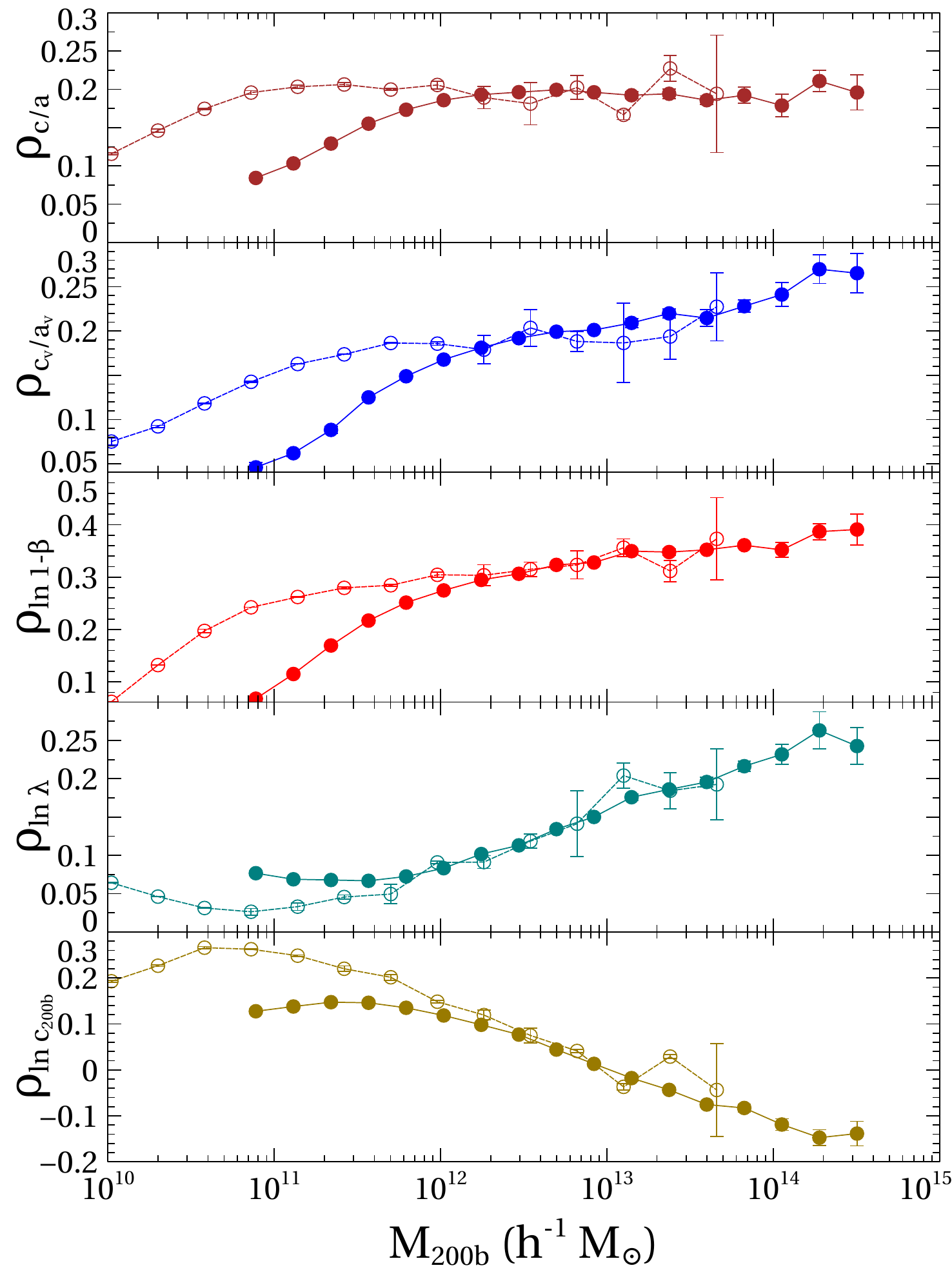}
 \caption{Similar to Figure~\ref{fig:convergencestudy0}, showing results for the Pearson correlation coefficient between halo property $c$ and standardised tidal anisotropy $\tilde{\alpha}$ (equation~\ref{eq:standardalpha}). }
    \label{fig:convergencestudy2}
\end{figure} 

In this Appendix, we study the numerical convergence of halo properties with dark matter particle count. In Figure~\ref{fig:convergencestudy0}, the median halo property is plotted as a function of mass for the high- and medium-resolution simulations. For each mass bin, the empty markers show the median halo property from haloes having $8\times$ more particles than the haloes shown with filled markers. Apart from relatively minor box size effects, any significant deviation between the trends seen in the two simulations especially at lower masses is due to resolution effects. 

As an example, in the top left panel which shows $\avg{c/a}$, there is significant degradation in the trend for the filled markers below $10^{12}\Mh$ which corresponds to a particle count  $\sim500$, below which the halo finder is incapable of correctly assessing the halo shape. Similar particle counts between $\sim 400$ (spin and concentration) and $\sim1000$ (velocity ellipsoid asphericity and velocity anisotropy) can be observed for other halo properties shown in the other panels. 
The median of the tidal anisotropy parameter $\avg{\ln\alpha}$ (bottom right panel), on the other hand, is well-converged between the two simulations even for haloes with only 30 particles in the medium-resolution box. 
Similar results are found in Figure~\ref{fig:convergencestudy1} for the numerical convergence of the standard deviation of halo properties. 

Figure~\ref{fig:convergencestudy2} shows the numerical convergence of the Pearson correlation between each halo property and the tidal anisotropy (see section~\ref{sec:propsfromenv} for a discussion of why the  Pearson coefficient is more appropriate for our analysis).  In general, with the exception of halo spin, we see that the effect of numerical convergence issues at low masses is to decrease the strength of these correlations and randomise the halo-environment dependence. Hence we expect numerical convergence issues to erase the assembly bias signal for haloes with poor resolution.

\section{Halo bias}

\subsection{Measuring halo bias in simulations}
\label{sec:measuringbias}
Our measurements of halo bias in narrow bins of halo mass follows \citet{2018MNRAS.476.3631P}. This is
is essentially equivalent to the ratio in Fourier space of the halo-matter cross-power spectrum to the matter auto-power spectrum, averaged over small $k$ values ($\leq0.1\ensuremath{h{\rm Mpc^{-1}}}$). In practice, we use a halo-centric measurement of halo bias \citep{2018MNRAS.476.3631P}, using the weights discussed by \citet{pa20}, whose arithmetic mean in narrow mass bins is identical to the above mentioned traditional estimate. 

\subsection{Comparison of Mocks with a higher resolution simulation}
\label{sec:comphighres}

\begin{figure}
\includegraphics[width=0.475\textwidth]{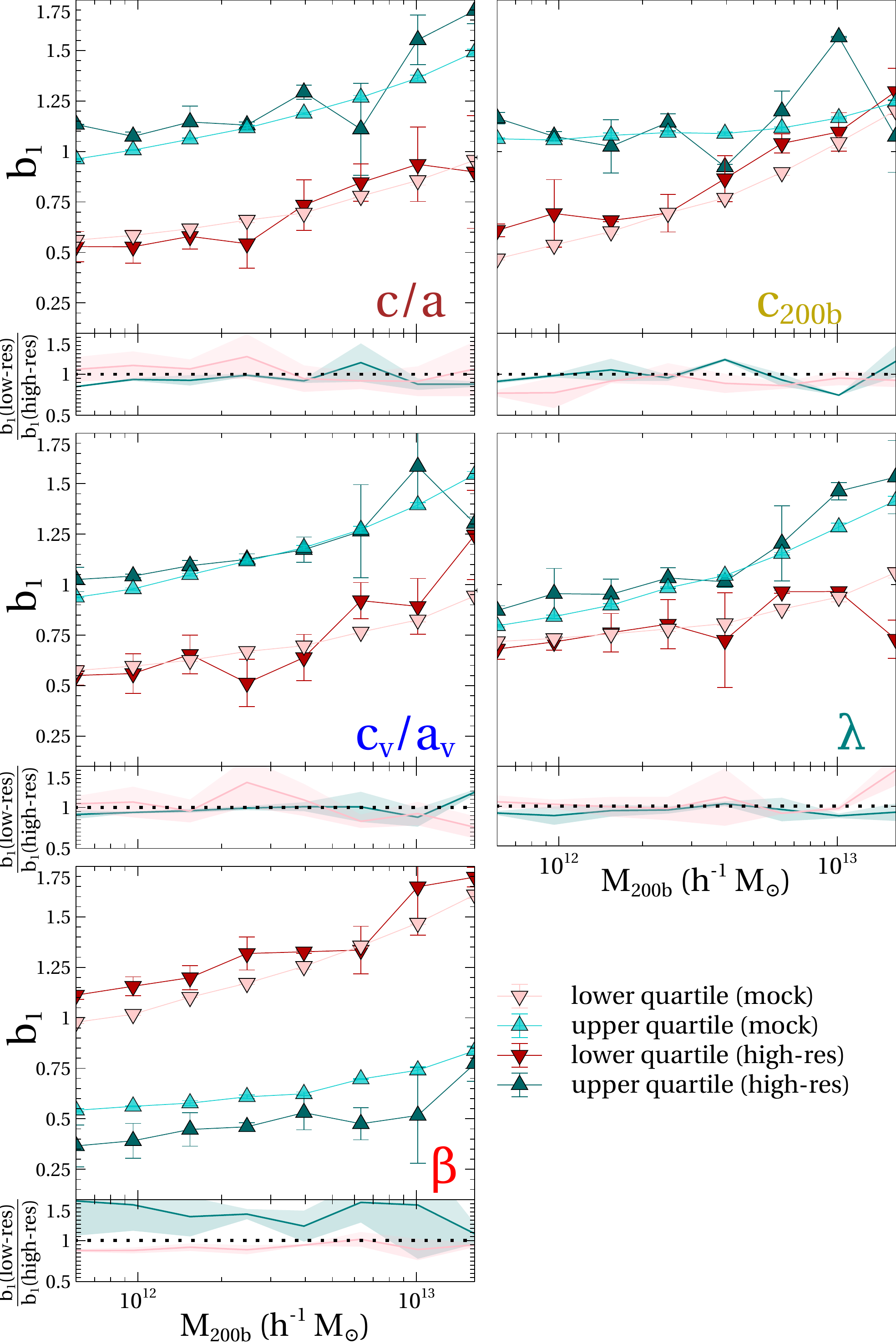} 
\caption{
Same as right panels of Figures~\ref{fig:largevolmocksu1} and~\ref{fig:largevolmocksu2}, but using as a reference the measurements of large-scale bias $b_1$ directly from the \emph{high-resolution} simulations (darker symbols). 
Note that in the mass range shown here, which corresponds to haloes having 30-700 CDM particles in the low-resolution simulation, we do not expect the assembly bias measured directly in the low-resolution simulation to be accurate due to numerical convergence issues, and we therefore do not display it. Our mock technique, on the other hand, agrees well with the high-resolution simulation.
}
\label{fig:largevolmock}
\label{lastpage}

\end{figure}

In this Appendix, we compare the assembly bias from the mock algorithm applied to the low-resolution simulations with that measured directly in the high-resolution simulation. Figure~\ref{fig:largevolmock} shows the results. The mass range which is displayed corresponds to haloes having (30-700) particles in the low-resolution simulation. We see good agreement for all the five halo properties. See  the main text for a discussion.

\end{document}

%% file: tables/c_to_a_mean.tex
\renewcommand{\arraystretch}{1.5} 
 \begin{tabular}{lllll}
 \hline 
 \hline  
&${(c/a)_0}$&${(c/a)_1}$&${(c/a)_2}$&$\chi^2(\rm 10\ d.o.f)$\\ \hline 
value&0.5591&-0.1686&0.0489&31.17\\ \hline 
std dev&0.0002&0.0015&0.0035 \\ \hline 
corr ${(c/a)_0}$&1.0000&0.0900&-0.3511 \\ \hline 
corr ${(c/a)_1}$&-&1.0000&-0.6860 \\ \hline 
 
 \hline \\ 
\end{tabular}

%% file: tables/c_to_a_sig.tex
\renewcommand{\arraystretch}{1.5} 
 \begin{tabular}{lllll}
 \hline 
 \hline  
&${\sigma^{c/a}_0}$&${\sigma^{c/a}_1}$&${\sigma^{c/a}_2}$&$\chi^2(\rm 10\ d.o.f)$\\ \hline 
value&0.1227&-0.0152&-0.0162&914.81\\ \hline 
std dev&0.0001&0.0004&0.0014 \\ \hline 
corr ${\sigma^{c/a}_0}$&1.0000&0.6399&-0.6814 \\ \hline 
corr ${\sigma^{c/a}_1}$&-&1.0000&-0.0144 \\ \hline 
 
 \hline \\ 
\end{tabular}

%% file: tables/c_to_a_corr.tex
\renewcommand{\arraystretch}{1.5} 
 \begin{tabular}{lll}
 \hline 
 \hline  
&${\rho^{c/a}_0}$&$\chi^2(\rm 12\ d.o.f)$\\ \hline 
value&0.1976&16.36\\ \hline 
std dev&0.0010 \\ \hline 
 
 \hline \\ 
\end{tabular}

%% file: tables/vc_to_va_mean.tex
\renewcommand{\arraystretch}{1.5} 
 \begin{tabular}{llll}
 \hline 
 \hline  
&${(c_{v}/a_{v})_0}$&${(c_{v}/a_{v})_1}$&$\chi^2(\rm 10\ d.o.f)$\\ \hline 
value&0.8153&-0.0775&302.95\\ \hline 
std dev&0.0001&0.0004 \\ \hline 
corr ${(c_{v}/a_{v})_0}$&1.0000&0.9969 \\ \hline 
 
 \hline \\ 
\end{tabular}

%% file: tables/vc_to_va_sig.tex
\renewcommand{\arraystretch}{1.5} 
 \begin{tabular}{lllll}
 \hline 
 \hline  
&${\sigma^{c_{v}/a_{v}}_0}$&${\sigma^{c_{v}/a_{v}}_1}$&${\sigma^{c_{v}/a_{v}}_2}$&$\chi^2(\rm 9\ d.o.f)$\\ \hline 
value&0.0689&0.0173&-0.0159&113.61\\ \hline 
std dev&0.0001&0.0009&0.0020 \\ \hline 
corr ${\sigma^{c_{v}/a_{v}}_0}$&1.0000&-0.4842&0.0908 \\ \hline 
corr ${\sigma^{c_{v}/a_{v}}_1}$&-&1.0000&-0.7460 \\ \hline 
 
 \hline \\ 
\end{tabular}

%% file: tables/vc_to_va_corr.tex
\renewcommand{\arraystretch}{1.5} 
 \begin{tabular}{llll}
 \hline 
 \hline  
&${\rho^{c_{v}/a_{v}}_0}$&${\rho^{c_{v}/a_{v}}_1}$&$\chi^2(\rm 11\ d.o.f)$\\ \hline 
value&0.1969&0.0683&14.31\\ \hline 
std dev&0.0009&0.0055 \\ \hline 
corr ${\rho^{c_{v}/a_{v}}_0}$&1.0000&-0.1106 \\ \hline 
 
 \hline \\ 
\end{tabular}

%% file: tables/beta_mean.tex
\renewcommand{\arraystretch}{1.5} 
 \begin{tabular}{llll}
 \hline 
 \hline  
&${\beta_0}$&${\beta_1}$&$\chi^2(\rm 11\ d.o.f)$\\ \hline 
value&-0.2353&-0.1484&144.95\\ \hline 
std dev&0.0003&0.0014 \\ \hline 
corr ${\beta_0}$&1.0000&-0.2186 \\ \hline 
 
 \hline \\ 
\end{tabular}

%% file: tables/beta_sig.tex
\renewcommand{\arraystretch}{1.5} 
 \begin{tabular}{llll}
 \hline 
 \hline  
&${\sigma^{\beta}_0}$&${\sigma^{\beta}_1}$&$\chi^2(\rm 11\ d.o.f)$\\ \hline 
value&0.1677&-0.0097&707.43\\ \hline 
std dev&0.0001&0.0005 \\ \hline 
corr ${\sigma^{\beta}_0}$&1.0000&0.9511 \\ \hline 
 
 \hline \\ 
\end{tabular}

%% file: tables/beta_corr.tex
\renewcommand{\arraystretch}{1.5} 
 \begin{tabular}{llll}
 \hline 
 \hline  
&${\rho^{\beta}_0}$&${\rho^{\beta}_1}$&$\chi^2(\rm 11\ d.o.f)$\\ \hline 
value&0.2480&0.1434&21.20\\ \hline 
std dev&0.0004&0.0051 \\ \hline 
corr ${\rho^{\beta}_0}$&1.0000&0.4235 \\ \hline 
 
 \hline \\ 
\end{tabular}

%% file: tables/spin_mean.tex
\renewcommand{\arraystretch}{1.5} 
 \begin{tabular}{lllll}
 \hline 
 \hline  
&${{\lambda}_0}$&${{\lambda}_1}$&${{\lambda}_2}$&$\chi^2(\rm 11\ d.o.f)$\\ \hline 
value&-3.3200&-0.1015&-0.1935&210.42\\ \hline 
std dev&0.0006&0.0023&0.0055 \\ \hline 
corr ${{\lambda}_0}$&1.0000&0.3336&-0.3470 \\ \hline 
corr ${{\lambda}_1}$&-&1.0000&0.7657 \\ \hline 
 
 \hline \\ 
\end{tabular}

%% file: tables/spin_sig.tex
\renewcommand{\arraystretch}{1.5} 
 \begin{tabular}{llll}
 \hline 
 \hline  
&${\sigma^{\lambda}_0}$&${\sigma^{\lambda}_1}$&$\chi^2(\rm 12\ d.o.f)$\\ \hline 
value&0.5862&0.0273&85.05\\ \hline 
std dev&0.0005&0.0021 \\ \hline 
corr ${\sigma^{\lambda}_0}$&1.0000&0.2233 \\ \hline 
 
 \hline \\ 
\end{tabular}

%% file: tables/spin_corr.tex
\renewcommand{\arraystretch}{1.5} 
 \begin{tabular}{llll}
 \hline 
 \hline  
&${\rho^{\lambda}_0}$&${\rho^{\lambda}_1}$&$\chi^2(\rm 12\ d.o.f)$\\ \hline 
value&0.0902&0.2060&22.52\\ \hline 
std dev&0.0010&0.0050 \\ \hline 
corr ${\rho^{\lambda}_0}$&1.0000&0.2810 \\ \hline 
 
 \hline \\ 
\end{tabular}

%% file: tables/c200b_corr.tex
\renewcommand{\arraystretch}{1.5} 
 \begin{tabular}{llllll}
 \hline 
 \hline  
&${\rho^{c_{200b}}_0}$&${\rho^{c_{200b}}_1}$&${\rho^{c_{200b}}_2}$&${\rho^{c_{200b}}_3}$&$\chi^2(\rm 10\ d.o.f)$\\ \hline 
value&0.1386&-0.5483&0.1734&0.1103&633.94\\ \hline 
std dev&0.0013&0.0083&0.0294&0.0444 \\ \hline 
corr ${\rho^{c_{200b}}_0}$&1.0000&0.5345&-0.4823&0.1381 \\ \hline 
corr ${\rho^{c_{200b}}_1}$&-&1.0000&0.1417&-0.5147 \\ \hline 
corr ${\rho^{c_{200b}}_2}$&-&-&1.0000&-0.8201 \\ \hline 
 
 \hline \\ 
\end{tabular}